# Room-Temperature Quantum Hall Effect in Graphene

K.S. Novoselov[1], Z. Jiang[2,3], Y. Zhang[2], S.V. Morozov[1], H.L. Stormer[2], U. Zeitler[4], J.C. Maan[4], G.S. Boebinger[3], P. Kim[2*] & A.K. Geim[1*]

[1]*Department of Physics, University of Manchester, M13 9PL, Manchester, UK*
[2]*Department of Physics, Columbia University, New York, New York 10027, USA*
[3]*National High Magnetic Field Laboratory, Tallahassee, Florida 32310, USA*
[4]*High Filed Magnet Laboratory, Radboud University Nijmegen, 6525 ED Nijmegen, Netherlands*

The quantum Hall effect (QHE), one example of a quantum phenomenon that occurs on a truly macroscopic scale, has been attracting intense interest since its discovery in 1980 (*1*). The QHE is exclusive to two-dimensional (2D) metals and has elucidated many important aspects of quantum physics and deepened our understanding of interacting systems. It has also led to the establishment of a new metrological standard, the resistance quantum $h/e^2$ that contains only fundamental constants of the electron charge $e$ and the Planck constant $h$ (*2*). As many other quantum phenomena, the observation of the QHE usually requires low temperatures $T$, typically below the boiling point of liquid helium (*1*). Efforts to extend the QHE temperature range by, for example, using semiconductors with small effective masses of charge carriers have so far failed to reach $T$ above 30K (*3,4*). These efforts are driven by both innate desire to observe apparently fragile quantum phenomena under ambient conditions and the pragmatic need to perform metrology at room or, at least, liquid-nitrogen temperatures. More robust quantum states, implied by their persistence to higher $T$, would also provide added freedom to investigate finer features of the QHE and, possibly, allow higher quantization accuracy (*2*). Here, we show that in graphene – a single layer of carbon atoms tightly packed in a honeycomb crystal lattice – the QHE can be observed even at room temperature. This is due to the highly unusual nature of charge carriers in graphene, which behave as massless relativistic particles (Dirac fermions) and move with little scattering under ambient conditions (*5*).

Figure 1 shows the room-$T$ QHE in graphene. The Hall conductivity $\sigma_{xy}$ reveals clear plateaux at $2e^2/h$ for both electrons and holes, while the longitudinal conductivity $\rho_{xx}$ approaches zero (<10Ω) exhibiting an activation energy $\Delta E \approx 600$K (Fig. 1B). The quantization in $\sigma_{xy}$ is exact within an experimental accuracy of $\approx 0.2\%$ (see Fig. 1C). The survival of the QHE to such high temperatures can be attributed to the large cyclotron gaps $\hbar\omega$ characteristic to Dirac fermions in graphene. Their energy quantization in magnetic field $B$ is described by $E_N = v_F \sqrt{|2e\hbar BN|}$ where $v_F \approx 10^6$ m/s is the Fermi velocity and $N$ an integer Landau level (LL) number (*5*). The expression yields an energy gap $\Delta E \approx 2800$K at $B = 45$T if the Fermi energy $E_F$ lies between the lowest Landau level $N=0$ and the first excited one $N=\pm 1$ (inset, Fig. 1B). This implies that, in our experiments at room temperature, $\hbar\omega$ exceeded the thermal energy $k_B T$ by a factor of 10. Importantly, in addition to the large cyclotron gap, there are a number of other factors that help the QHE in graphene to survive to so high temperatures. First, graphene devices allow for very high carrier concentrations (up to $10^{13}$ cm$^{-2}$) with only a single 2D subband occupied, which is essential to fully populate the lowest LL even in ultra-high $B$. This is in contrast to traditional 2D systems (for example, GaAs heterostructures) which are either depopulated already in moderate $B$ or exhibit multiple subband occupation leading to the reduction of the effective energy gap to values well below $\hbar\omega$. Second, the mobility $\mu$ of Dirac fermions in our samples does not change appreciably from liquid-helium to room temperature. It remains at $\approx 10,000$ cm$^2$/Vs, which yields a scattering time of $\tau \sim 10^{-13}$ sec so that the high field limit $\omega\tau = \mu \cdot B \gg 1$ is reached in fields of several T.

These characteristics of graphene foster hopes for the room-$T$ QHE observable in fields significantly smaller than 30T. In fact, we observe the Hall plateaus developing already in $B < 20$T at 300K. The need for high $B$ is attributed to broadened LLs due to disorder, which reduces the activation energy. We expect that improving sample homogeneity and achieving higher $\mu$ (currently limited by static defects) should allow the observation of the room-$T$ QHE using conventional magnets. This should open up new vistas for developing graphene-based resistance standards (certainly, operational above liquid-nitrogen temperature) and for novel quantum devices working at elevated temperatures.

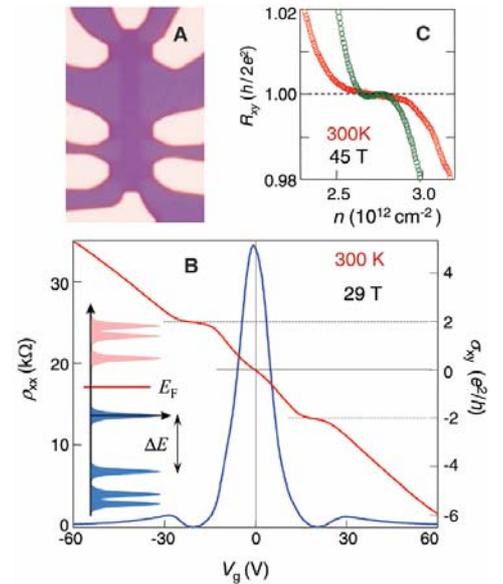

Figure 1. **Room-temperature QHE in graphene.** (**A**) – Optical micrograph of one of the devices used in the measurements. The scale is given by the Hall bar's width of 2 μm. Device fabrication procedures were described in (*5*). (**B**) - $\sigma_{xy}$ (red) and $\rho_{xx}$ (blue) as a function of gate voltages $V_g$ in a magnetic field of 29 T. Positive (negative) gate voltages $V_g$ induce electrons (holes) in concentrations $n = (7.2 \cdot 10^{10}$ cm$^{-2}$/V$) \cdot V_g$ (*5*). The inset illustrates the Landau level quantization for Dirac fermions. (**C**) - Hall resistance $R_{xy}$ for electrons (red) and holes (green) shows the accuracy of the observed quantization at 45 T.

*e-mail: pkim@phys.columbia.edu and geim@man.ac.uk

[1] S. Das Sarma and A. Pinczuk, *Perspectives in Quantum Hall Effects*, Wiley, New York (1997).
[2] B. Jeckelmann, B. Jeanneret. *Rep. Prog. Phys.* **64**, 1603 (2001).
[3] S.Q. Murphy *et al. Physica E* **6**, 293 (2000).
[4] G. Landwehr *et al. Physica E* **6**, 713 (2000).
[5] K.S. Novoselov *et al. Nature* **438**, 197 (2005); Y. Zhang, Y.W. Tan, H.L. Stormer, P. Kim. *Nature* **438**, 201 (2005).

[6] This work was supported by EPSRC (UK), Royal Society and Leverhulme Trust, NSF (DMR-03-52738), DOE (DE-AIO2-04ER46133, DE-FG02-05ER4615), FOM (Netherlands), Microsoft Corporation, and W. M. Keck Foundation. The experiments were partially performed at the National High Magnetic Field Laboratory (supported by NSF Cooperative Agreement # DMR-0084173, by the state of Florida, and by DOE) and partially at the European High Field Magnet Laboratory (Nijmegen).